\documentclass[showpacs,prl,twocolumn,floatfix]{revtex4}

\usepackage{graphicx}
\usepackage{hyperref}
\usepackage[dvipsnames,usenames]{color}

\begin{document}

\title{The simplicity of planar networks}

\author{Matheus P. Viana$^1$, Emanuele Strano$^2$, Patricia Bordin$^{3,4}$, Marc Barthelemy$^{5.6}$}

\affiliation{$^1$ Department of Developmental and Cell Biology, University of
California, Irvine, CA 92697, USA.}

\affiliation{$^2$ LaSIG, Geographic Information Systems Laboratory
EPFL, Ecole Polytechnique F\'ed\'erale de Lausanne
Station 18, CH-1015 Lausanne, CH}

\affiliation{$^3$ Universit\'e Paris-Est, Institut de Recherche en
  Constructibilit\'e, ESTP, F-94230, Cachan, France}

\affiliation{$^4$ Universit\'e Paris Diderot, Sorbonne Paris Cit\'e, Institut des Energies de Demain (IED), 75205 Paris, France}

\affiliation{$^5$ Institut de Physique Th\'{e}orique, CEA, CNRS-URA 2306, F-91191, 
Gif-sur-Yvette, France}

\affiliation{$^6$ Centre d'Analyse et de Math\'ematiques Sociales, EHESS,
  190-198 avenue de France, 75244 Paris}

\begin{abstract}
  Shortest paths are not always simple. In planar networks, they can
  be very different from those with the smallest number of turns - the
  simplest paths. The statistical comparison of the lengths of the
  shortest and simplest paths provides a non trivial and non local
  information about the spatial organization of these graphs. We
  define the simplicity index as the average ratio of these lengths
  and the simplicity profile characterizes the simplicity at different
  scales. We measure these metrics on artificial (roads, highways,
  railways) and natural networks (leaves, slime mould, insect wings)
  and show that there are fundamental differences in the organization
  of urban and biological systems, related to their function,
  navigation or distribution: straight lines are organized
  hierarchically in biological cases, and have random lengths and
  locations in urban systems. In the case of time evolving networks,
  the simplicity is able to reveal important structural changes during
  their evolution.
\end{abstract}

\maketitle

A planar network is a graph that can be drawn on the
two-dimensional plane such that no edges cross each other
\cite{Clark:1991}. Planar graphs pervade many aspects of science: they
are the subject of numerous studies in graph theory, in combinatorics
\cite{Tutte:1963,Bouttier:2004} and in quantum gravity
\cite{Ambjorn:1997}. Planar graphs are also central in biology where
they can be used to describe veination patterns of leaves or insect
wings. In particular, the vasculature of leaves
\cite{Weitz:2012,Katifori:2012} displays an interesting architecture
with many loops at different scales, while in insects, the vascular
network brings strength and flexibility to their wings. In city science,
planar networks are extensively used to represent, to a good
approximation, various infrastructure networks
\cite{Barthelemy:2011}. In particular, transportation networks
\cite{Haggett:1969} and more recently streets patterns
\cite{Hillier:1984,Marshall:2006} are the subject of many studies
\cite{Jiang:2004,Roswall:2005,Porta:2006,Porta:2006b,Lammer:2006,Crucitti:2006,Cardillo:2006,Gastner:2006,Xie:2007,Jiang:2007,Masucci:2009,Chan:2011,Courtat:2011,Strano:2012,Strano:2013,Barthelemy:2013}
that are trying to characterize both topological (degree distribution,
clustering, etc.) and geometrical (angles, segment length, face area
distribution, etc.) aspects of these networks.

Despite a large number of studies on planar networks, there is still a
lack of global, high-level metrics allowing to characterize their
structure and geometrical patterns. Such a characterization is however
difficult to achieve and in this article, we will discuss an important
aspect of planar graphs which is intimately connected to their
geometrical organization. In this respect, we will define new metrics
and test them on various datasets, both artificial (roads, highways,
railways, and supply networks) and natural (veination patterns of
leaves and wings, slime mould) enabling us to obtain new information
about the structure of these networks.

\begin{figure*}[ht!]
\includegraphics[width=\linewidth]{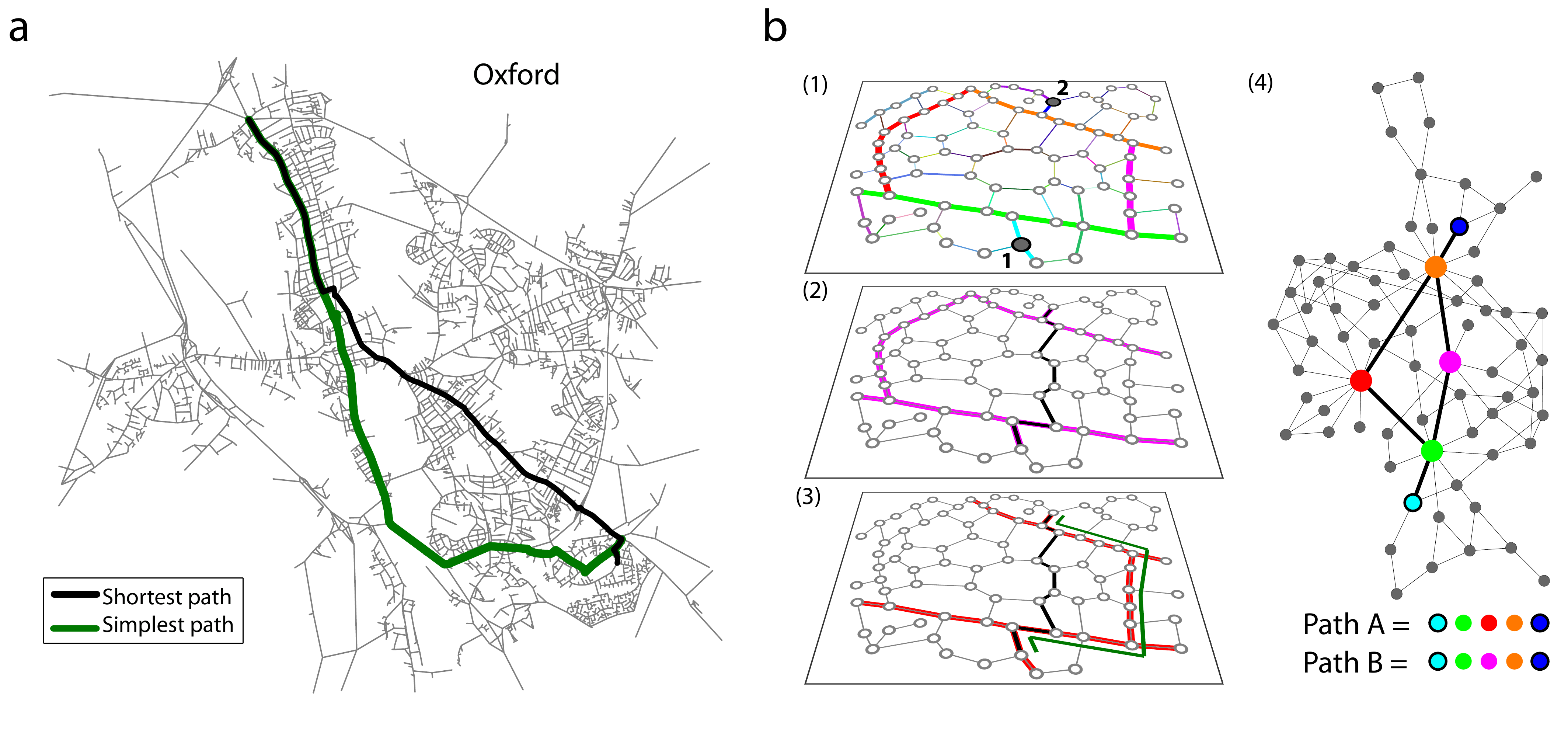}
\caption{{\bf Simplest path: example and calculation}. (a) Example of
  shortest (black line) and simplest (dark green line) paths
  illustration on the Oxford (UK) street network. The simplest path
  has less turns at the expense of being longer than the shortest
  path. (b) In (1) we show a planar network and in (4) its dual
  representation. The colors of straight lines in (1) corresponds to
  the ones of nodes in (4). The simplest path between nodes 1 and 2 is
  obtained by the shortest path in the dual space (between nodes cyan
  and blue in this case).  There are two paths $A$ and $B$ with length
  $4$ (corresponding to magenta lines in (2) and to red lines in (3)),
  and the shortest one is chosen as the simplest path (green line in
  (3)). For comparison we also show the shortest path (black line in
  (2,3)). The figure was created using NetworkX.}
\label{fig:illus}
\end{figure*}

We will now introduce the main metrics used in this article. Generally
speaking, we can define different types of paths for a given pair of
nodes $(i,j)$. A usual quantity is the shortest euclidean path of
length $\ell(i,j)$ which minimizes the distance travelled to go from
$i$ to $j$. We can however ask for another path which minimizes the
number of turns - the simplest path, of length $\ell^*(i,j)$ (if there
are more than one such path we choose the shortest one).
Fig.~\ref{fig:illus}a displays an example of the shortest and simplest
path for a given pair of nodes on the Oxford (UK) street network.

To identify the simplest path, we first convert the graph from the
primal to the dual representation, where each node corresponds to a
straight line in the primal graph. These straight lines are determined
by a continuity negotiation-like algorithm, as described in Material
and Methods.  Edges in dual space, in turn, represent the intersection
of straight lines in the primal graph (see Fig.~\ref{fig:illus}b).

We define the number of turns $\tau$ of a given path as the number of
switches from one straight line to another when walking along this
path. This quantity is intimately related to the amount of information
required to move along the path \cite{Roswall:2005}. We have computed
the probability distribution $P(\tau)$ for all shortest and simplest
paths of several networks (see SI) and the results show that this
distribution is usually centered around a smaller values for the
simplest paths than for the shortest paths, as expected. More
generally, we show in the Supplementary Information that the average
number of turns $\langle\tau\rangle$ versus the number of nodes $N$
indeed displays a small-world type behavior characterized by a slow
logarithmic increase with $N$, consistently with previous analysis of
the dual network \cite{Roswall:2005,Porta:2006b}. This feature is thus
not very useful to distinguish different networks and shows that the
distribution of the number of turns is a very partial information and
tells very little about the spatial structure of the simplest
paths. For navigation purposes (neglecting all congestion effects) and
in order to understand the structure of the network, it is useful to
compare the lengths of the shortest and the simplest paths with the
ratio $\ell^*(i,j)/\ell(i,j)\geq 1$. It is then natural to introduce
the {\it simplicity index} $S$ as the average
\begin{equation}
S=\frac{1}{N(N-1)}\sum_{i\neq j}\frac{\ell^*(i,j)}{\ell(i,j)}.
\end{equation}
The simplicity index is larger than one and exactly equal to one for a
regular square lattice and any tree-like network for example. Large
values of $S$ indicate that the simplest paths are on average much
longer than the shortest ones, and that the network is not easily
navigable. We note here that we do not take into account congestion
effects which can influence the path choice (see for example
\cite{Yan:2006}). This new metric is a first indication about the spatial
structure of simplest paths but mixes various scales, and in order to
obtain a more detailed information, we define the {\it simplicity
  profile}
\begin{equation}
S(d)=\frac{1}{N(d)}\sum_{i,j/d_E(i,j)=d}\frac{\ell^*(i,j)}{\ell(i,j)},
\end{equation}
where $d_E(i,j)$ is the euclidean distance between $i$ and $j$ and
where $N(d)$ is the number of pairs of nodes at euclidean distance
$d$.  This quantity $S(d)$ is larger than one and its variation with
$d$ informs us about the large scale structure of these graphs. We can
draw a generic shape of this profile: for small $d$, we are the scale
of nearest neighbors and there is a large probability that the
simplest and shortest paths have the same length, yielding $S(d\to
0)\sim 1$, and increasing for small $d$. For very large $d$, it is
almost always beneficial to take long straight lines when they exist, thus reducing
the difference between the simplest and the shortest paths. As a
result we expect $S(d)$ to decrease when $d\to d_{max}$ (note that a
similar behavior is observed for another quantity, the route-length
efficiency, introduced in \cite{Aldous:2010}). The simplicity profile
will then display in general at least one maximum at an intermediate scale $d^*$
for which the length differences between the shortest and the simplest
path is maximum.  The length $d^*$ thus represents the typical size of
domains not crossed by long straight lines. At this intermediate
scale, the detour needed to find long straight lines for the simplest
paths is very large.

We finally note here that these indices are actually not limited to
planar networks but to all networks for which the notion of straight
lines has a sense and can be computed. This would be the case for
example for spatial networks which are not perfectly
planar~\cite{Barthelemy:2011}.

We introduce a null model in order to provide a simple benchmark to
further analyze the results obtained by these new metrics (the
expression 'null model' should be understood here in the sense of the
benchmark and not in the usual statistical definition). The goal in
this study is to compare empirical results with a very simple model
based on a minimal number of assumptions, but we note that it would be
also interesting to compare various models generating planar
networks. We start with $N$ points randomly distributed in the plane
and construct the Voronoi graph (see the Supplementary Information for
further details). We then add a tunable number of straight lines of
length $\ell$ distributed according to
$P(\ell)\sim\ell^{-\alpha}$. Examples of networks generated by this
model as well as many results are shown in the SI.

\begin{figure*}[ht!]
\includegraphics[width=\linewidth]{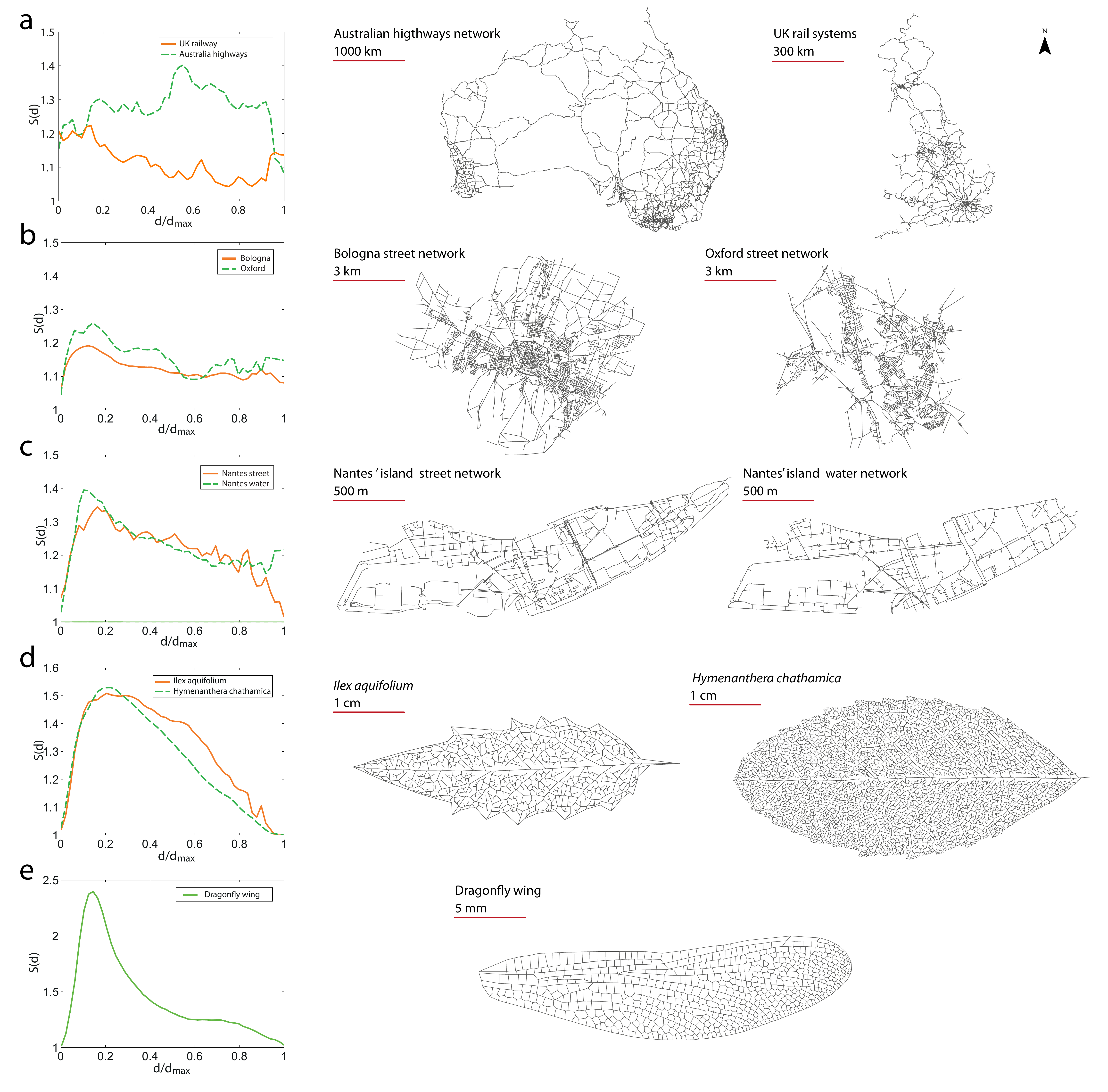}
\caption{{\bf Simplicity profiles}. We represent here the simplicity
  profiles for different networks ranging from large
  scale networks ($10^6m^2$) to small scales of order
  $10^{-3}m$. We see on these different examples the effect of the
  presence of long straight lines and of a polycentric structure. In
  particular for cases (d,e), we can clearly see that the peak at
  $d^*\sim 0.2d_{max}$ corresponds to the size of domains not crossed
  by long straight lines. The figure was created using ESRI ArchMap
  10.1 and Adobe Illustrator.}
\label{fig:all}
\end{figure*}

We first study static networks (see Fig. \ref{fig:all}) such as the streets of cities
(Bologna, Italy; Oxford, UK; Nantes, France), the national highway network of
Australia, the national UK railway system, and the water supply
network of central Nantes (France). In the case of biological networks, we study the
veination patterns of leaves ({\it Ilex aquifolium} and {\it Hymenanthera
chatamica}), and of a dragongly wing. Details on these datasets can be
found in the SI.

\begin{figure*}[ht!]
\includegraphics[width=\linewidth]{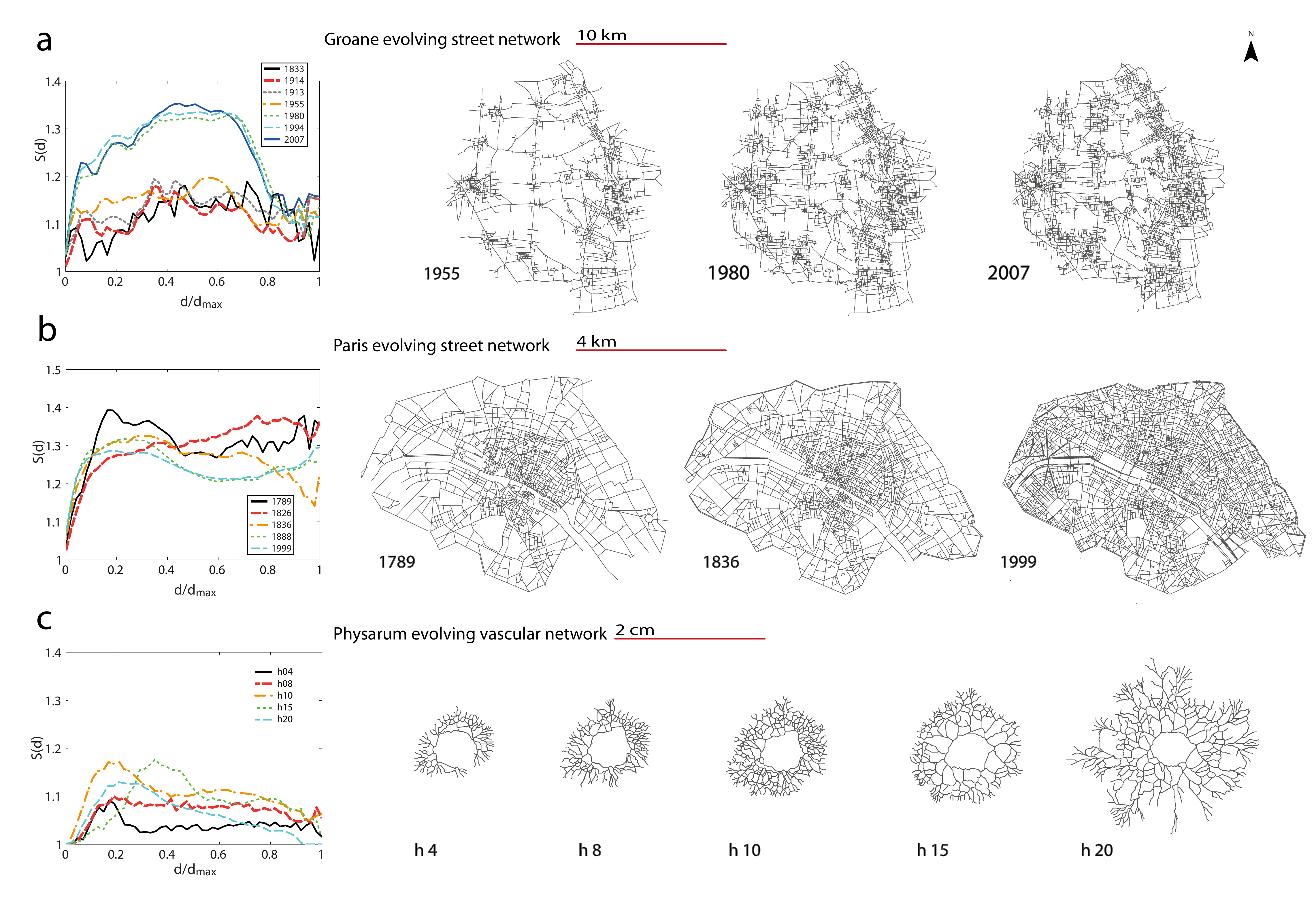}
\caption{{\bf Simplicity profiles for time-varying networks}. We
  represent here the profiles for (a) the road network of the Groane
  region (Italy), (b) the street network of Paris (France) in the
  pre-Haussmannian (1789, 1836) and post-Haussmannian (1999) periods,
  and in (c) the {\it Physarum} network growing on a period of one day
  approximately. We observe on (a) and (b) that the evolution of the
  profile is able to reveal important structural changes. In (c) the
  evolution follows closely the one obtained with the null model (see
  SI). The figure was created using ESRI ArchMap
  10.1 and Adobe Illustrator.}
\label{fig:evolution}
\end{figure*}

We also consider three datasets describing the time evolution of
networks at different scales (see Fig.~\ref{fig:evolution}): at a
small scale and in the biological realm we study the evolution of a
slime mould network. At the city scale we present results on the road
network of Paris (France) from 1789 until now. Paris was largely
transformed by a central authority (the prefect Haussmann under
Napoleon III) in the middle of the 19th century and the dataset
studied here displays the network before and after these important
transformations, offering the possibility to study quantitatively the
effect of top-down planning \cite{Barthelemy:2013}. At the multi-town
level, we study the road network of the Groane area (Italy) (see the
SI for details on these datasets). These networks allow us to explore
different systems at very different scales from $10^{-3}$ (Slime
mould) to $10^6$ (Australian highways) meters.

\begin{figure*}[ht!]
\includegraphics[width=\linewidth]{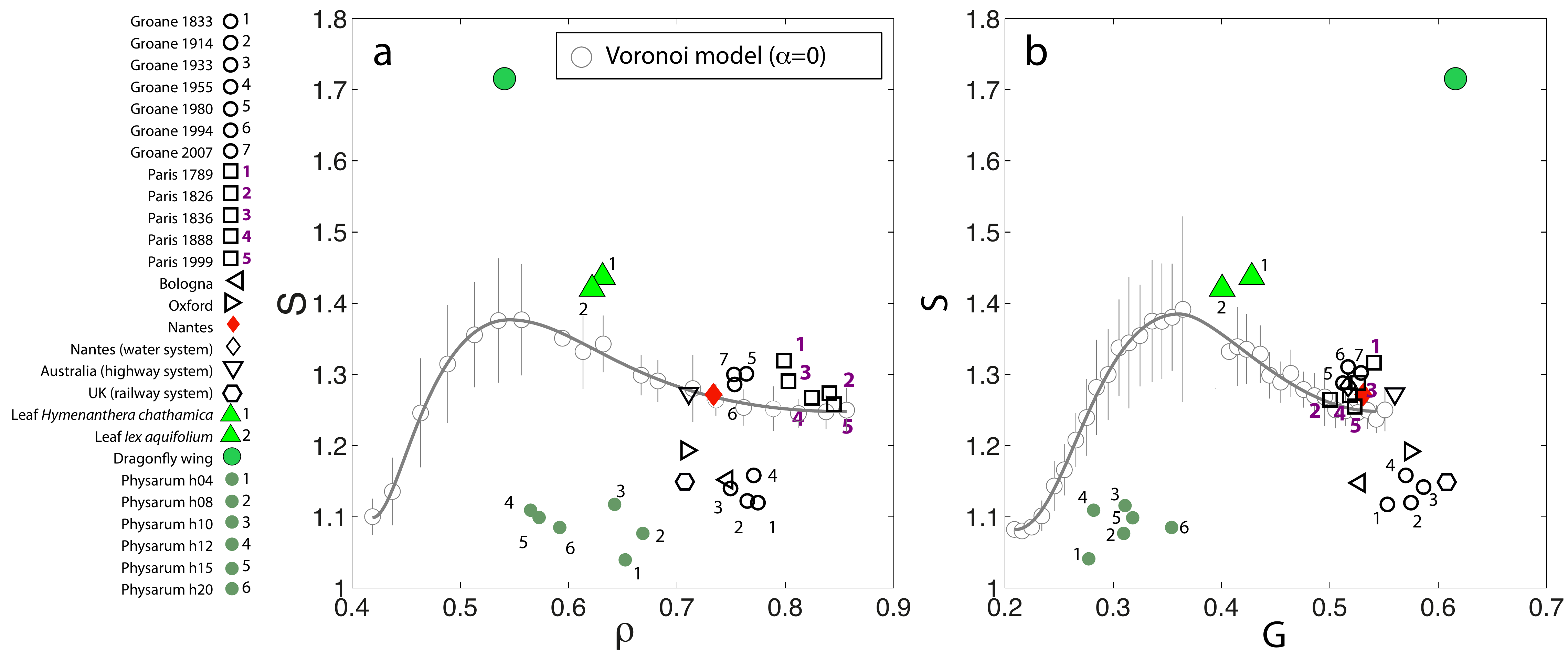}
\caption{{\bf Simplicity index}.  Simplicity versus (a) the density of straight line $\rho$
  and (b) the Gini coefficient for the length of straight
  lines. In both plots, the symbols correspond to the different
  networks studied here. We also represented the result for the null
  model (for $\alpha=0$) and its cubic spline interpolation
  (continuous line). From (a) we see that biological networks are
  limited to the region $\rho\leq 0.7$ and have a large simplicity
  index, and from (b) we see that urban networks have simultaneously
  higher values of $G$ and relatively small values of $S$.}
\label{fig:Svalues}
\end{figure*}

We compute the simplicity index $S$ for the various
datasets and for the null model as well. The results are shown in
Fig. \ref{fig:Svalues} as a function of the density of straight lines $\rho$ and
the Gini coefficient $G$ for the length of straight lines (see
Material and Methods for details). The density $\rho$ of straight
lines is defined as the ratio of total length of straight lines (see Fig. SI2),
over the total length of the network, and $G$ is an indicator
of the diversity of the length of straight lines.

The first observation from Fig.~\ref{fig:Svalues} is that the
simplicity index encodes information which is neither contained in the
density $\rho$ nor in the Gini coefficient $G$, and reveals how the
straight lines are distributed in space and participate in the flows
on the network.

In Fig.~\ref{fig:Svalues}a, we observe that the density of straight lines is
always larger for urban systems. More precisely, in the biological
systems the density lies in the range $\rho\in [0.55,0.7]$, while we
observe $\rho>0.7$ for artificial systems. Except for the Physarum,
which appears to be close to a regular lattice with a small simplicity
and small Gini coefficient, the simplicity index for the wing and the
leaves is larger than the values obtained for the null model. These
results indicate that the organization of straight lines in biological systems
is very different from artificial systems, that have very similar values
of $\rho$, $G$, and $S$. In particular, we observe a hierarchy 
of straight lines in biological systems (see Fig~\ref{fig:all}): a main artery
(the midrib for leaves) connects to veins which in turn are connect
to smaller veins and so on. In the case of dragonfly wing, the main
straight line is given by the external border of the network. The
existence of these main straight lines in biological systems will impact
the structure of simplest paths and impose some large detour, resulting
in a larger value of the simplicity index. 

For urban systems, the simplicity is very close to the null model (of
order $1.3$ in this density range), suggesting that in dense urban
systems, long straight lines are added at random (An exception
concerns, the pre-Haussmannian Paris (1789-1836) for which we observe
a simplicity smaller than for the null model, the reason being
probably that the networks at these times were very sparse). As a
result, navigation on urban systems requires relatively less
information with no additional cost: the simplest path is not too
different from the shortest path. 

Finally, we note an interesting effect in the null model in
Fig.~\ref{fig:Svalues}a which is the existence of a maximum of the
simplicity at densities of order $\rho\sim0.55$. In this density
regime, using straight lines implies having to make large detours. However,
when the density exceeds $0.6$, there are enough straight lines to
enable a simplest path which differs not too much from the shortest
one.


We now discuss the simplicity profile shown in Fig.~\ref{fig:all}.  We
observe that basically for most of these systems, the simplicity
profile displays the generic shape with a maximum at an intermediate
scale. In urban cases, such as Bologna and central Nantes, we have a
typical monocentric system with a dense center and a few important
radial straight lines, leading to a simple profile $S(d)$. In the case
of Oxford and the Australian highway network, the polycentric
organization leads to multiple peaks in the simplicity profile
(Fig.~\ref{fig:all}). Interestingly, we observe that the profiles for
australian highways and railways in the UK are very different, despite
their similar scale, density $\rho$, and Gini coefficient $G$. In
particular, the UK railway displays small values of the simplicity
(less than $\lesssim1.2$) while for the Australian highway network
there are many pairs of nodes for which the simplest path is much
longer than the shortest one. We also observe that the profile for
both street and water systems of Nantes have a very similar shape,
pointing to the fact that these networks are strongly correlated. In
addition, the position and the height of the peak ($\approx 1.4$)
observed for the Nantes water system suggests that this distribution
system has similar features compared to biological systems such as
vein networks in leaves (see below) whose function is also
distribution.

Compared to urban systems, the simplicity profile of biological
networks have a single well-defined, and much more pronounced peak. We
observe values of order $S_{max}\approx 1.5$ and $2.5$ for
$d^*/d_{max}\approx 0.2$, meaning that for this range of distance, the
detour made by the simplest path is very large. This peak is related
to the existence of domains of typical size $d^*$ not crossed by large
veins. We see here a clear effect of the existence of the spatial
organization of long straight lines in these systems, probably
optimized for the distribution (of water for leaves). The decay for
large $d$ is also much faster in the biological case compared to urban
systems: this shows that in biological systems there are long straight
lines allowing to connect far away nodes. This is particularly evident
on the leaves shown in Fig.~\ref{fig:all} where we can see the first
levels (primary and secondary) veins, the rest forming a network. For
streets, the organization is much less rigid and the hierarchy less
strict: we have a more uniform spatial distribution of straight lines,
leading to a smoother decrease of $S(d)$.


Going beyond static networks, we apply our new metrics to study the
structural changes of time evolving networks.

The first example of a time evolving network is the road network of
the Groane region which is a $125km^2$ area located north of Milan
\cite{Strano:2012}. We have 7 snapshots of this network for different
times from 1833 to 2007 (see the Supplementary Information for
details). This region evolved without central planning and is thus a
good example of an `organic' evolution of urban
systems. The simplicity profile shown in Fig.~\ref{fig:evolution}(a)
allows us to distinguish two different periods. The first period from
1833 to 1955 displays a relatively small simplicity at all scales,
while a distinct second regime appears from 1980 until now. In this
latter regime, the simplicity profile is substantially larger for all
scales. This is an effect of the massive urban densification, leading
to a polycentric structure where the readability and the ease to
navigate are drastically lowered.

At a smaller scale, we study the evolution of central Paris between
1789 and 1999. This dataset provides an interesting case study, as
Paris experienced larges changes due to Haussmann in the middle of the
19th century (see \cite{Barthelemy:2013} for details and more
references about this network). This is an opportunity to observe
quantitatively the effect of top-down planning: until 1836, we are in
the pre-Haussmann Paris, while from 1888 until now we are in the
post-Haussmann period. The effect of Haussmann's central planning is
clearly visible on the network shown in
Fig.~\ref{fig:evolution}(b). From 1789 to 1836, we have a relatively
large simplicity at all scales and we observe a decrease in that
period at small scales ($d/d_{max}\lesssim 0.4$) which corresponds well to
the fact that many religious and aristocratic domains and properties
were sold and divided in order to create new houses and new roads,
improving congestion inside Paris. The 1826-1836 transition displays a
decrease of the simplicity for distance larger than roughly $5$ kms
(corresponding to $d/d_{max}\approx 0.6$) indicating that long
distance routes were simplified. It is interesting to note that
during this period the eastern part of Paris experienced large
transformations with the construction of the channel
St. Martin. Finally in the period 1836 to 1888, when Paris experienced
Haussmann's transformation, the simplicity profile is strongly
affected: compared to 1836, the simplicity is improved in the range
$d/d_{max}\in [0.3,0.8]$, which can be attributed to the construction of large avenues
connecting important nodes of the city. In addition, we observe the
surprising effect that at large scales $d/d_{max}\gtrsim 0.8$, the simplicity is
degraded by Haussmann's work: this however could be an artifact of the
method and the fact that we considered a portion of Paris only and
neglected the effect of surroundings.

Finally, we note that differences between Groane and Paris might be
explained in terms of a sparse, polycentric urban settlement
(Groane) versus a dense one (Paris). In particular, in the
`urban' phase for Groane (after 1955), the simplicity profile becomes
similar to the one of a dense urban area such as Paris.

Finally, we show the results in Fig.~\ref{fig:evolution}(c) for
the {\it Physarum Policephalum}, a biological system evolving at the
centimeter scale. {\it Physarum} is a unicellular multinucleated
amoeboid that during its vegetative state takes a complex shape. Its
plasmodium viscous body whose goal is to find and connect to food
sources, crystallizes in a planar network-like structure of
micro-tubes \cite{Nakagaki07112004}.  In simple terms, Physarum's
foraging strategy can be summarized in two phases: i) the exploration
phase in which it grows and reacts to the environment and ii) the
crystallization phase in which it connects to food sources with micro-tubes. We
inoculated active plasmodium over a single food source and observe the
micro-tube network at six phases of its growth (see SI for
details). Under these conditions, we observe that the network is
statistically isotropic around the food source as shown in
Fig.~\ref{fig:evolution}(c) and develops essentially radially. We
first observe that the simplicity profile for the Physarum is
relatively low (less than $\lesssim1.2$), suggesting that
simplicity could be an important factor in the evolution of this organism. A
closer observation shows that during its evolution, the Physarum adds
new links to the previous network and also modifies the network on a
larger scale, as revealed by the changes of the simplicity
profile. The evolution of the profile is similar to the one obtained
for the null model when the density is increased (see SI), suggesting
that the statistics of straight lines in this case could be described as
essentially resulting from the random addition of straight lines of random
lengths (with $\alpha=2$).


We have shown that the new metrics introduced here encode in a useful
way both topological and geometrical information about the global
structure of planar graphs. In particular, our results highlight the
structural differences between biological and artificial networks. In
the former, we have a clear spatial organization of straight lines,
with a clear hierarchy of lines (midrib, veins, etc), leading to
simplest paths that require a very small number of turns but at the cost of large
detours. In contrast, there is no such strong spatial organization in
urban systems, where the simplicity is usually smaller and comparable
to a null model with straight lines of random length and
location. These differences between biological and urban systems might
be related to the different functions of these networks: biological
networks are mainly distribution networks serving the purpose of
providing important fluids and materials. In contrast, the role of
road networks is not only to distribute goods but to enable
individuals to move from one point of the city to another. In
addition, while biological networks are usually the result of a single
process, urban systems are the product of a more complex evolution
corresponding to different needs and technologies.

These new metrics also allow us to track important
structural changes of these networks. The simplicity profile
thus appears as a useful tool which could provide a quantitative
classification of planar graphs and could help in constructing a
typology of leaves or street patterns for example.

\section*{Methods}

\subsection*{Measuring the simplest path}
All the simplest paths of a given network were calculated in the dual
space by converting the networks from the primal to the dual
representation, where straight lines are mapped into nodes and the
intersection between straight lines were mapped into edges. Straight
lines are found by using a version of the {\it ICN} (Intersection
Continuity Negotiation) algorithm \cite{Porta:2006b}. More
specifically, given an edge $(i,j)$, we search among the adjacent
edges attached to $j$, $(j,k)$, that one that is most aligned to
$(i,j)$. If the angle $\theta_{i,j,k}$ between $(i,j)$ and $(j,k)$ is
smaller or equal to $\theta_c=30^\circ$, we assume that these two edges belong
to the same straight line. This procedure continues until no more
edges are assigned to the same straight line. Then, the procedure is
repeated in opposite direction starting from the adjacent edges
attached to node $i$. Once assigned to a straight line, an edge is
removed from the network. As it is, this algorithm produces different networks depending on the
choice for the initial edge. To overcome this ambiguity, our algorithm
always starts with the edge that give us the longest straight line for
a given network. After this straight line is fully detected and its
edges deleted, we choose the next edge that will give us the second
longest straight line and so on. The algorithm ends when there are no
more edges left in the network.

Once all straight lines have been identified, the dual representation
is built by looking at the intersection between straight
lines. Each straight line is mapped onto a node in the dual
space and two nodes are connected together if their respective
straight lines intersect each other at least once. To illustrate this
process, we show in Fig~\ref{fig:illus}(b,1) an example of planar network in the primal
representation where the edges are colored according the $id$ of the
straight line they belong to. In Fig. \ref{fig:illus}(b,4) we show the dual
representation of the same network. It is important to note that the
longest straight lines, in this example represented by orange, red,
green and magenta give rise to hubs in the dual space.

In order to calculate the simplest path between nodes 1 and 2 from
Fig. \ref{fig:illus}(b,1), we search for the shortest path between their
respective straight lines in the dual space, cyan and blue in this
case. As it can be seen, there are two paths with length 4, $A$ and
$B$. Each of them define a subgraph in the primal representation, here
represented by the set of magenta lines in Fig.~\ref{fig:illus}(b,2) and
red lines in Fig.~\ref{fig:illus}(b,3) for paths A and B,
respectively. Then we evaluate the shortest path distance between
nodes 1 and 2 over these subgraphs and we adopted the shortest one as
the simplest path - green dashed in Fig. \ref{fig:illus}(b,3). The black path in Figs. \ref{fig:illus}(c,d) represent
the shortest path between nodes 1 and 2.

\subsection*{Gini coefficient}
The Gini coefficient quantifies the inequalities of the lengths of
straight lines, and is defined as in~\cite{Gini:1987}
\begin{equation}
\label{eq:gini}
G_k = \frac{1}{2 E^2 \bar{\ell}} \sum_{i,j=1}^{E} | \ell_i - \ell_j |
\end{equation}
where $\bar{\ell}$ is the average length of straight lines and $E$ is
the number of straight lines. The Gini coefficient lies in the range
$[0,1]$ and $G=0$ when all lengths are equal. On the
other hand, if all lengths but one are very small, the Gini
coefficient will be close to $1$. 

\section*{Acknowledgments} 

  We thank Prof. A. Perna for the dataset of the leaf {\it
    Hymenanthera chathamica} and Prof. A. Adamatzky for providing us
  the {\it Physarum Policephalum}. MB thanks H. Berestycki and
  M. Gribaudi for interesting discussion. MB acknowledges funding from
  the EU commission through project EUNOIA (FP7-DG.Connect-318367).

\section*{Author contributions}

MPV, ES, PB, MB designed, performed research and wrote the paper.

\section*{Additional information}
The authors declare no competing financial interests.

\end{document}